\documentclass[11pt]{article}
\usepackage[utf8]{inputenc}
\usepackage{amsmath}
\usepackage{graphicx}
\usepackage{algorithmicx}
\usepackage{algpseudocode}
\usepackage{algorithm}

\usepackage{authblk}

\renewcommand{\P}{\mathbf{P}}

\usepackage{natbib}

\begin{document}
\title{Point process models for sweat gland activation observed with noise}

\author[1]{Mikko Kuronen}

\author[1]{Mari Myllymäki}

\author[2]{Adam Loavenbruck}

\author[3]{Aila Särkkä}

\affil[1]{Natural Resources Institute Finland (Luke)}

\affil[2]{Department of Neurology, Kennedy Laboratory, University of Minnesota}

\affil[3]{Mathematical Sciences, Chalmers University of Technology and the University of Gothenburg}

\date{}

\maketitle

\abstract{

The aim of the paper is to construct spatial models for the activation of sweat glands for healthy subjects and subjects suffering from peripheral neuropathy by using videos of sweating recorded from the subjects. The sweat patterns are regarded as realizations of spatial point processes and two point process models for the sweat gland activation and two methods for inference are proposed. Several image analysis steps are needed to extract the point patterns from the videos and some incorrectly identified sweat gland locations may be present in the data. To take into account the errors we either include an error term in the point process model or use an estimation procedure that is robust with respect to the errors.

}

\section{Introduction}

Assessment of sudomotor (sweat) function has long been used in clinical and research settings for detection of neurologic dysfunction \citep{CoonEtal1941, LaderEtal1962}. Minor’s starch iodine test was originally described in 1928 \citep{Minor1928}, where  tincture of iodine was applied to the skin, air dried, and then powdered with corn starch. Sweating is stimulated with increasing room temperature or use of pilocarpine. As sweat flows from pores, iodine is diluted and the solution absorbed by the starch powder, turning dark black from yellow. Normally the entire skin surface should be able to sweat in response to sufficient stimuli, and absence of sweating in an area of the body is indicative of loss of neurologic function in that area.  Sweating is critical in human evolution in maintaining ability to thermoregulate in a wide range of climates and activity levels. Neurologic control, headquartered in the hypothalamis, is therefore tightly regulated and concordantly disruptive in pathologic states such as peripheral neuropathy.

Peripheral neuropathy is a disease state of peripheral nerves, the segment of the nervous system which extends from the brain and spinal cord to various targets in the body, such as muscles, sensory receptors and autonomically controlled organs. Peripheral neuropathy occurs in etiologically diverse conditions which cause damage or dysgenesis of peripheral nerves. The most common causes include diabetes, toxicity such as in chemotherapy and excessive alcohol consumption, and vitamin deficiencies  \citep{LoavenbruckEtal2017}.  The resulting nerve damage causes various combinations of muscle weakness, pain, numbness and autonomic dysfunction.

Autonomic nerves are often the earliest to be affected in peripheral neuropathy \citep{Sumner2003, Said2007, SharmaEtal2015}, and penetrate all parts of the body, including digestive tract, liver, kidneys, bladder, genitals, lungs, pupils, heart, and skin.  Skin is the largest organ in the body, and contains a vast network of the distal ends of sensory and autonomic nerves over the entire body surface. These distal ends of nerves are especially susceptible to systemic disease. Because sweating is neutrally controlled and modulated, and can be measured at the skin surface, it can be used to reflect alterations in the underlying nerves.

Currently there are several tests used in clinical practice to evaluate sudomotor function \citep{HilzEtal2006,IlligensEtal2009,MinotaEtal2019}.
Thermoregulatory sweat testing \citep{FealeyEtal1989} measures percentage of body surface area sweating elicited by heated, humidified sauna.  
Sweat imprints and silastic molds \citep{PapanasEtal2005,HarrisEtal1972}, measure the density and distribution of activated sweat glands in an area of skin.
Quantitative sudomotor axon reflex testing (QSART) is likely the most widely used autonomic test of sweating \citep{LowEtal1983,LowEtal2006}, comparing against robust normative databases the total volume of sweat produced by 1 $\text{cm}^2$ areas of skin at standardized sites.

\begin{figure}
\centering
\includegraphics[width=\textwidth]{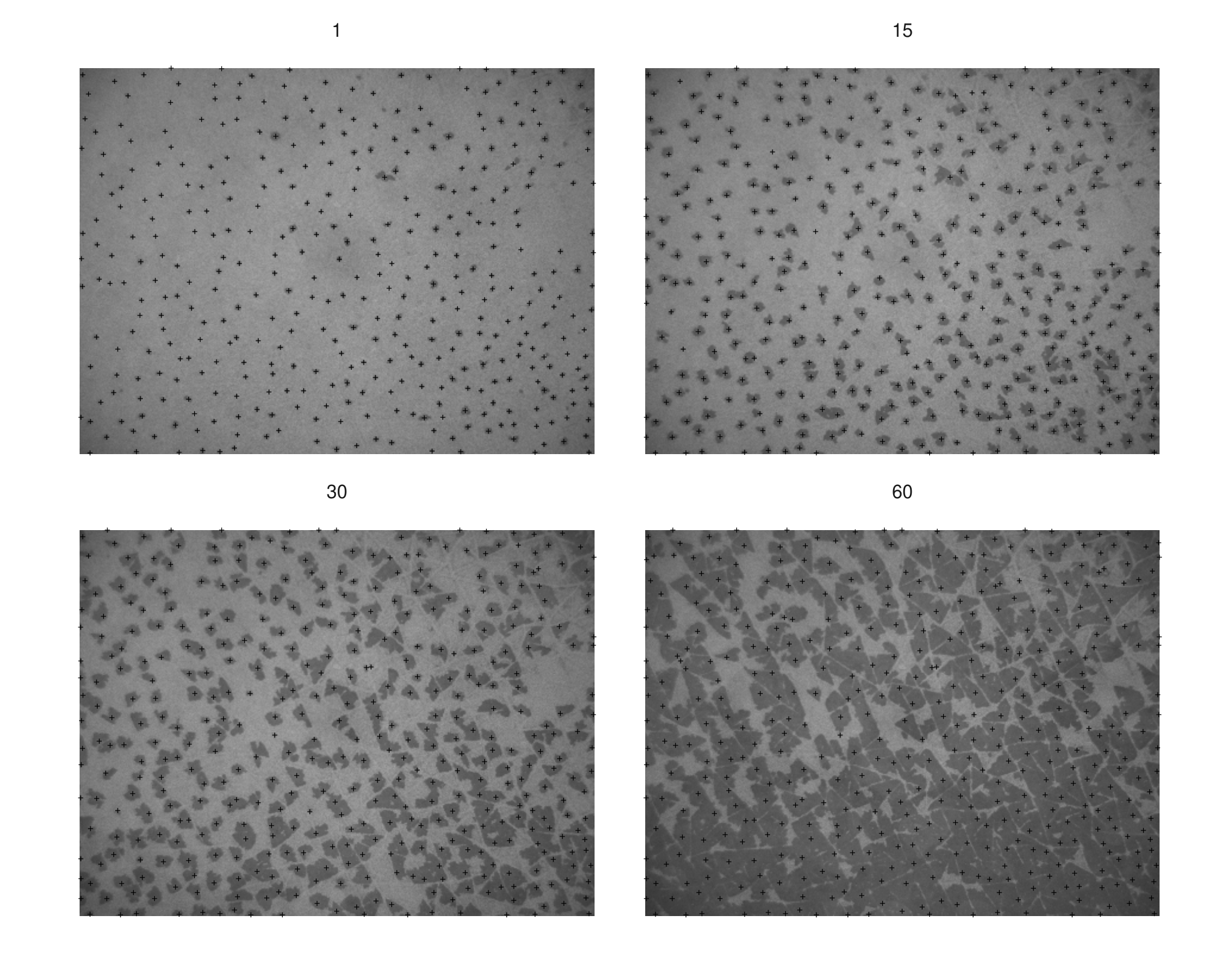}
\caption{A sequence of snapshots (1 sec (top left), 15 sec (top right), 30 sec (bottom left), and 60 sec (bottom right)) of one control subject with extracted gland locations (+).}
\label{fig:4frames}
\end{figure}

The sensitive sweat test (SST) enhanced Minor’s starch iodine test with closeup time lapse imaging, and software analysis \citep{ProviteraEtal2010,KennedyEtal2013,LoavenbruckEtal2017,LoavenbruckEtal2019}. The critical feature of the test is a rigid, transparent video screen which limits sweat to an essentially two-dimensional space. As sweat exits ducts, it dilutes the iodine painted on the skin onto starch coated plastic film. The imaged result is a field of sharply demarcated, dark sweat spots on a white background, expanding centripetally around the opening of each duct (Figure \ref{fig:4frames}). The area of each spot is therefore a measurement of the volume of sweat produced by each gland. Sub-nanoliter volumes of liquid were measured by pipette and shown to create reproducible sweat spot areas. Similarly, tracking the increase in sweat spots’ areas between timelapse frames measures the rate of sweating from each duct at the nanoliter level. Of note, blackened areas of film do not return to white during the test – sweat spots can only enlarge. The videos therefore provide several measurable physiologic data points, including coordinates and relative locations of all sweat spots, the second by second volumes of sweat (nanoliters) and flow rate of each sweat gland (nL/minute), total number of activated sweat glands, density of activated sweat glands (glands/$\text{cm}^2$), total sweat volume (nL), and total sweat rate (nL/minute).

Using the dynamic sweat test, a significant reduction of sweating was observed in diabetic subjects in the distal leg but not in forearm \citep{ProviteraEtal2010}. The study included measurements taken from the forearm of 14 diabetic subjects and 14 age- and sex-matched healthy controls and from distal leg of 7 diabetes patients and 7 controls.
In a larger study \cite{LoavenbruckEtal2017}, 178 healthy controls and 20 neuropathy subjects were tested, most of them at the hand, thigh, calf, and foot, some only at calf and foot, and it was concluded that neuropathy subjects had lower sweat rates per sweat gland, lower total amount of sweat, and lower sweat gland density. 
It was also observed visually that the sweat patterns of the diabetic subjects were less regular than the healthy patterns\cite{ProviteraEtal2010}. This visual inspection indicates that the spatial sweat patterns that the videos provide may reveal some additional abnormalities that may appear when the sweat rate and the total amount of sweat are still within normal ranges. However, up to now, no spatial analysis of the sweat patterns to quantify this observation has been performed.

In this paper, we concentrate on spatial analysis of the sweat patterns, regarding the locations of sweat spots or glands as realizations of spatial point processes. Our main emphasis is to develop suitable methodology for analysing the spatial structure of the sweat gland patterns and activation extracted from video sequences.
As the data are non-standard in point process literature, some special treatment is needed.

To extract the coordinates of the individual sweat spots, i.e.\ the points of the point patterns, from the videos (see Figure \ref{fig:4frames}), several image analysis steps are needed. As a non-standard step, we introduce an algorithm based on the detection of a change point in each pixel. This pixel-by-pixel approach suits to the video sequences, where the sweat does not dry once it has appeared, much better than going through the videos frame by frame, because the sweat gland locations are best detected at times where sweat first appears. However, even though we perform careful analysis of the videos, there are some spots that are incorrectly recorded as two spots due to, e.g.,\ wrinkles in the skin. 
It is not straightforward to remove these errors automatically and doing it manually can be very time consuming. Therefore, they need to be taken into account in the analysis of the point patterns.

Some studies of point patterns observed with errors or noise can be found in the literature. For example, in the area of minefield detection, one first detects a minefield and then classifies each observed point in the minefield either as mine or as noise. The observation window is typically divided into two parts, the minefield as a region with a higher intensity containing both mines and noise and a low intensity area containing only noise \cite{ByersRaftery1998}. The points can then be classified in a Bayesian set-up where a posterior probability for each point being a mine is derived \citep{CressieLawson2000,WalshRaftery2002}. In a similar Bayesian framework, classification of points of a superposition of a Strauss process and Poisson noise has been considered \citep{RedenbachEtal2015,RajalaEtal2016}. A likelihood of an imperfect observation given the true point process, where the imperfect observation can be due to random thinning,
displacement, censoring of the displaced points or superposition of additional points\cite{LundRudemo2000} and Bayesian analysis for similar data \cite{LundPenttinenRudemo1999} can also be found in the literature. Furthermore, a Bayesian framework for estimating the intensity of a noisy point process, where the noise is either due to points within the sampling window but regarded as being outside and/or points outside the window that were incorrectly regarded as points inside the window, is available \cite{ChakrabortyGelfand2010} as well as  descriptive statistics for noisy spatial point processes, where the noise is perturbation of points\cite{BarHenEtal2013}.

Here, we suggest two different ways to model the activation of sweat glands and to take noise into account in the analysis, by either including an error term in the model or using an estimation procedure that is robust with respect to the errors. We pay special attention to incorrectly recorded close pairs of points since they can cause problems for the analysis of regular point patterns such as our data.

We propose two models for the sweat gland patterns which are different in their nature.
In the first model, the activation of individual sweat glands is modelled by a sequential point process, where sweat spots appear conditionally on the pattern so far. The other model is more physiologically motivated, a generative point process model, where the activated sweat gland pattern is modelled as a thinning of the underlying true (unobserved) sweat gland pattern which is modelled first. While the likelihood of the sequential model is tractable, it has been considered computationally costly to evaluate \citep{PenttinenYlitalo2016}. Here, we propose an efficient way to perform traditional likelihood-based inference for a certain type of sequential models, which makes also likelihood-based Bayesian inference feasible. 
The likelihood of the generative model is not easily tractable and, therefore, we employ approximate Bayesian computation (ABC) to estimate the model parameters.
When some noise points are present, the sequential model is replaced by a mixture model having the sequential point process and an error point process as its components. In the generative model, the summary statistics in the ABC approach are chosen such that they are robust with respect to the errors. 

The rest of the paper is organized as follows. We first describe the extraction of the points from the videos and the preliminary analysis of the data in Section \ref{sec:prelim}. Then, we present the sequential and generative models together with a description of the inference methods in Sections \ref{sec:seqmodel} and \ref{sec:genmodel}, respectively. Further, the methodology is illustrated by analysing a set of video sequences taken from the right foot of 15 subjects, either controls or subjects with suspected or diagnosed neuropathy. The models are fitted separately to each subject. Section \ref{sec:discussion} is left for further discussion.
We carry out all computations in Julia \citep{BezansonEtal2017} while we use R
\citep{R2018} mainly for plotting.

\section{Data and preliminary data analysis}\label{sec:prelim}

\subsection{Description of data}
The data have been collected by Dr.\ Kennedy's group at the University of Minnesota by using the dynamic sweat test they have presented \citep{ProviteraEtal2010}. First, sweating is stimulated by placing a patch with pilocarpine gels on the test site, foot or calf. Then, the test site is dried and painted with iodine solution. Finally, a camera is placed on the skin and a video is recorded at the rate of 1 frame/sec for 60 seconds. The size of the frame was $2592\times 1944$ pixels corresponding to $17.5\times13 \text{ mm}^2$. Videos were taken from the feet and/or calves from 121 healthy controls without known neuropathy or known risk factors for neuropathy, as well as 72 subjects who had reported having symptoms of neuropathy, 20 of whom had well characterized neuropathy (diagnosis based on neurologic examination and nerve conduction studies). Therefore, the subjects were divided into three groups: controls, subjects with suspected neuropathy (MNA), and subjects with diagnosed neuropathy (MNA Diagnosed).

In this study, we have access to five videos from the right foot from each of the three groups. Based on earlier studies, it was clear that the number of activated sweat glands is an important predictor for the condition, controls having higher density than subjects in the neuropathy groups. To study whether the spatial features of the activated sweat gland patterns could indicate differences between the healthy and neuropathic subjects having similar densities, the five videos were selected based on the point density of the pattern so that different groups have overlapping densities.
A sequence of snapshots (1 sec, 15 sec, 30 sec, and 60 sec) of one control subject is shown in Figure \ref{fig:4frames}.
Here, we study the patterns of activated sweat glands at the end of videos
as realisations of spatial point processes. The complete video is needed to extract the gland locations, because these can be obtained most precisely at their first occurrence (see Section \ref{video-processing}).

\subsection{Video processing with change point detection}\label{video-processing}

Extracting the locations and sizes of the sweat spots required several video processing steps:
transforming the video into sweat/not sweat binary video, splitting the sweat part of the video into the sweat produced by individual sweat glands and finally extracting the point pattern of gland locations.

\begin{figure}
\centering
\includegraphics[width=\textwidth]{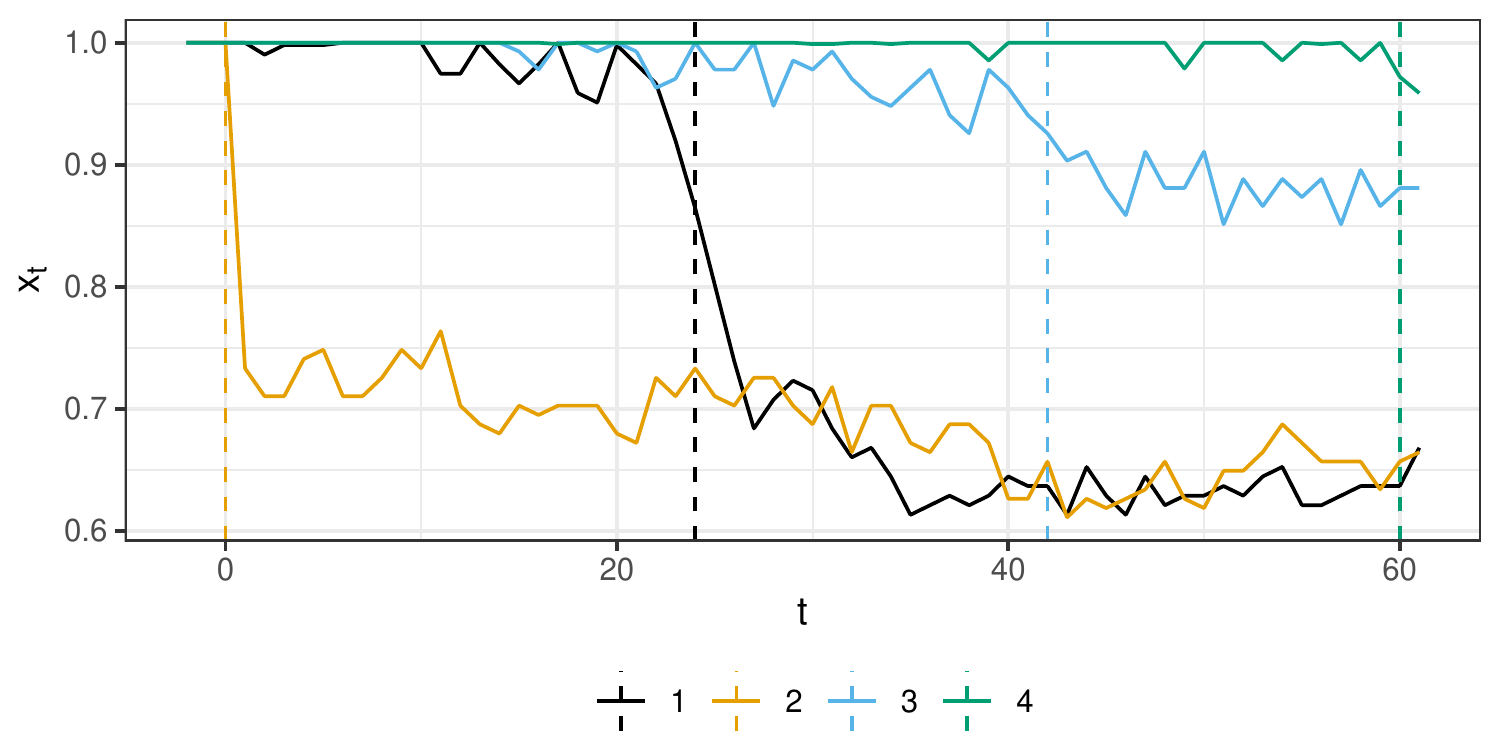}
\caption{Time series for four pixels with estimated jump locations (frames) marked by dashed lines.}\label{fig:jumps}
\end{figure}

The first step consisted of a background correction, finding change points, and finding and applying a threshold to the change points.
The background correction was done by kernel smoothing using a Gaussian kernel with $\sigma = 100$ pixels to the first frame of the video. Since the first frame had only small amount of sweat, the resulting image mostly mimicked the lighting conditions. For example, the corners of the frame were darker than the middle.
Next, the grayscale values $g_t$ of each pixel at frame $t$ were divided by the estimated lightning intensity $l$ of the pixel and the time series of these scaled grayscale values were considered to find the pixels that belong to the wet area. More precisely, a time series was constructed for each pixel as follows: Let $x_{-2} = x_{-1} = x_0 = 1$ and $x_t = g_t/l$ for $t = 1, \dots, T$. 
The change point of the time series $x_{-2}, x_{-1}, x_0, x_1, \dots, x_T$ was defined as the integer value $t \ge 1$ that minimizes
$f(t) = s_{-2,t}^2 + s_{t+1, T}^2$, where
\begin{equation*}
    s_{j,k}^2 = \frac{1}{k-j+1}\sum_{i=j}^k x_i^2 - \left(\frac{1}{k-j+1}\sum_{i=j}^k x_i\right)^2
\end{equation*}
is the sample variance of $x_j, \dots, x_k$. The time series and estimated change points for four pixels are shown in Figure \ref{fig:jumps}. Since each pixel, even the ones that do not belong to the wet area, obtained a change point, thresholding on the difference of sample means was used to filter out small changes. A per video threshold was found by trial and error evaluating the point patterns that resulted from the whole video processing visually. In Figure \ref{fig:jumps}, the two largest jumps, 1 and 2, passed the threshold. 
The resulting binary video frames were post processed with a morphological closing to fill in some small gaps.

The sweat area in the first frame was segmented into the sweat produced by individual glands. Starting with the second frame, each new sweat pixel was assigned to the closest spot in the previous frame. The distance was measured as the shortest path through the new sweat area. Several filtering steps were applied in various stages of the process to account for pixels that belonged to spots that were too small to be sweat.

\begin{figure}
\centering
\includegraphics[width=\textwidth]{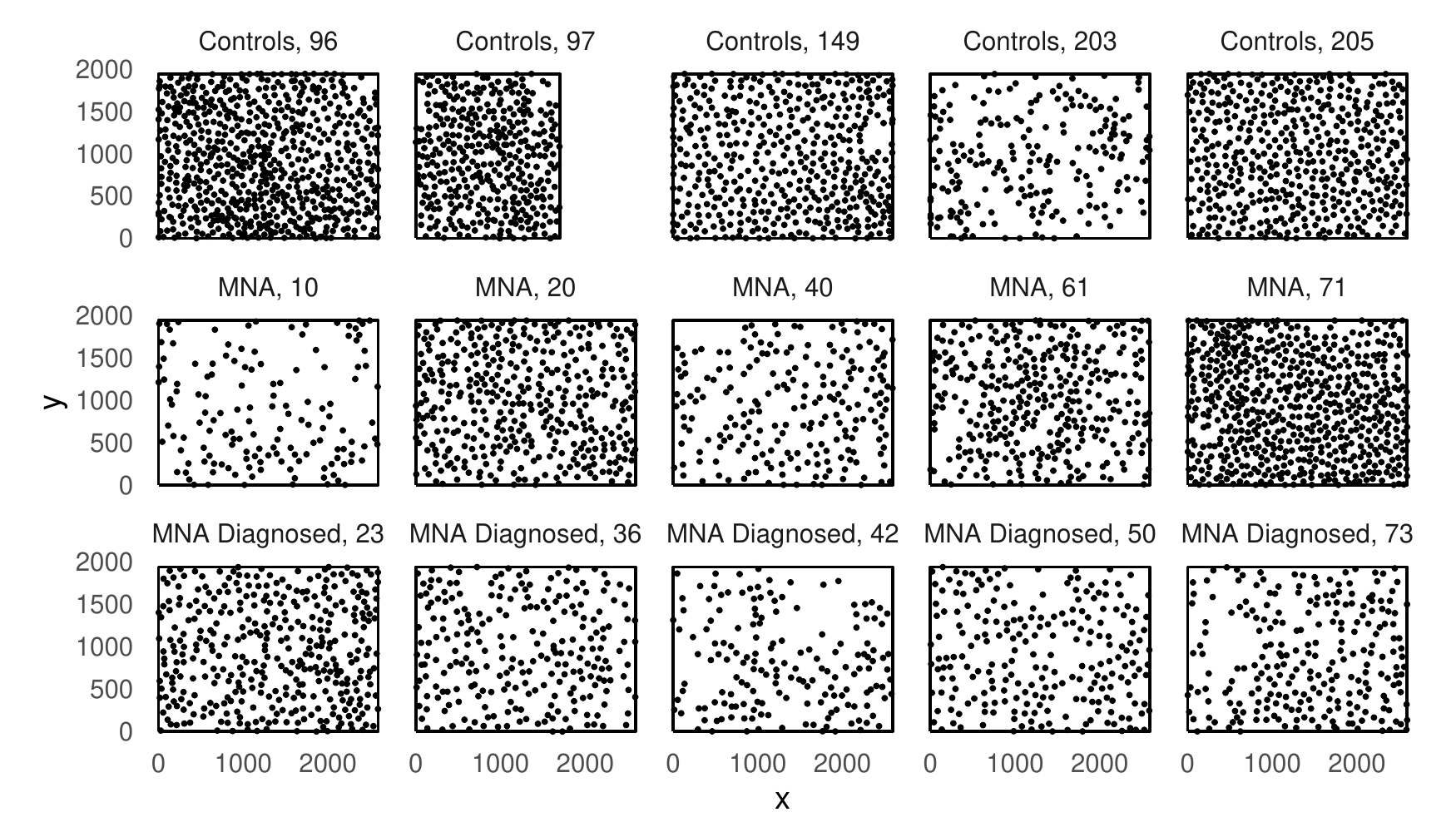}
\caption{Point patterns extracted from the videos for control subjects (top) and subjects with suspected (middle) and diagnosed (bottom) neuropathy.}
\label{fig:ppdata}
\end{figure}

Finally, we extracted a point pattern with coordinates for each gland. To obtain an ordered point pattern we used the frame of the first appearance, and for those spots that arrived at the same frame we used spot size as a surrogate for the time, where larger ones were assumed to have appeared earlier. 
An example of extracted point patterns in the video can be seen in Figure \ref{fig:4frames}.
Figure \ref{fig:ppdata} shows the final point patterns of all subjects.

\subsection{Spatial summary functions}

To analyse the spatial structure of the activated sweat gland patterns, we used two different commonly used spatial summary functions, the pair correlation function $g$ and the empty space function $F$. If the underlying point process is stationary and isotropic, these summary functions are functions of distance only. 

The pair correlation function $g$ summarizes the second-order property of point patterns\cite{IllianEtal2008}. Heuristically, $\lambda {\rm d}x g(r)$, where $\lambda$ is the intensity of the point process, is the probability that there is a point in an infinitesimal region with size ${\rm d}x$ at distance $r$ from an arbitrary point of the process.
To estimate the pair correlation function we used a traditional kernel smoothing method \citep{StoyanStoyan1994} with translational edge correction, the recommended bandwidth $0.15/\sqrt{\hat \lambda}$, where $\hat \lambda$ is the intensity estimated from the point pattern, and the Epanechnikov kernel.

The empty space function $F(r)$ measures the probability that an arbitrary location has the nearest point of the process within radius $r$.
It was estimated using Kaplan-Meier method \citep{BaddeleyGill1997}.

\subsection{Descriptive statistics of the point patterns}

\begin{figure}
\centering
\includegraphics[width=0.7\textwidth]{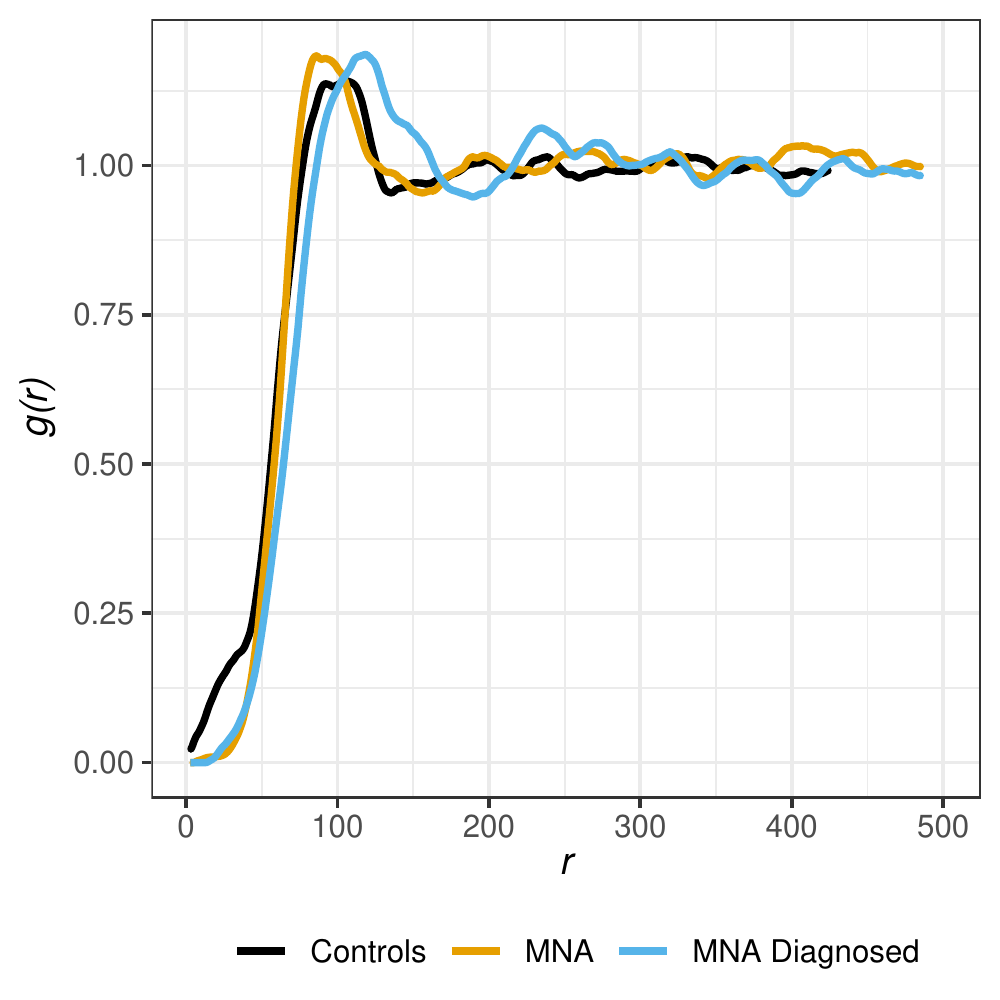}
\caption{Pooled pair correlation functions for the three groups.}
\label{fig:pooled_pcf}
\end{figure}

We first estimated the pair correlation function for each of the sweat gland patterns shown in Figure \ref{fig:ppdata} and thereafter, obtained the groupwise pair correlation functions (see Figure \ref{fig:pooled_pcf}) by pooling the estimated pair correlation functions of all the subjects within the group\citep{IllianEtal2008}. The individual pair correlation functions were weighted by the squared number of points when pooling. 
The pair correlation functions show a clear sign of inhibition in all three groups ($g(r)<1$ for small $r$).
Further, the first top of the functions appears approximately at the same distance for the control and suspected neuropathy groups.
However, the diagnosed neuropathy group has the first top at a slightly longer distance, indicating somewhat larger range of inhibition than in the other two groups.

At very short distances, especially the control subjects seem to have some unexpected close pairs of points.
Upon a closer inspection of the point patterns and the videos it was reasonable to assume that some  sweat spots had been detected as two nearby spots, instead of having merged into one.
An obvious, simple idea to remove such close pairs of spots would be to merge all small glands having a larger spot closer than at some specified distance with the larger spot. However, such erroneous pairs of glands may appear at various (small) distances from each other and thus, applying a global limiting distance is not reasonable. 
Instead of using an additional image analysis step, we include some of this inaccuracy in the modelling and/or parameter estimation.

\section{Sequential point process model}\label{sec:seqmodel}

Since sweat glands activate at different times, we modelled the activation by using sequential point processes similar to those suggested for modelling eye movements\citep{PenttinenYlitalo2016}. The points, activated sweat glands in our case, are generated sequentially conditioning on the already existing points. Points are added until the observed number of points in the pattern has been reached and the main focus here is to make inference on the arrival density.
Below, we first recall the general sequential model \citep{PenttinenYlitalo2016} (Section \ref{sec:seqmodel_general}) and specify it in our case without (Section \ref{sec:softcore-model}) and with noise (\ref{sec:mixture-model}). Further, we discuss efficient inference for the sequential models (Section \ref{sec:seqmodel_inf}) and, finally, fit the sequential model with noise to the sweat gland data (Section \ref{sec:seqmodel_data}).

\subsection{General sequential point process model}\label{sec:seqmodel_general}

Denote by $W$ the observation window and by $n$ the fixed number of points in the pattern. 
The first point $x_1$ is assumed to be uniformly distributed in $W$ and
the $k$th point, $k=2,\dots,n$, is assumed to follow
the density $y \mapsto f(y; \vec{\bf x}_{k-1})$, where
$\vec{\bf x}_{k-1} = (x_1, x_2, \dots, x_{k-1})$.
The density function for the whole point pattern $(x_1,...,x_N)$ is then
\begin{align*}
    \vec{\bf x}_n \mapsto \frac{1}{|W|} \prod_{k=2}^n f(x_k; \vec{\bf x}_{k-1}),
\end{align*}
where $1/|W|$ is the contribution of the first point.
A nice feature of the sequential point process models is that they have a tractable likelihood even though it can be costly to compute \citep{PenttinenYlitalo2016}.

\subsection{Soft-core model}\label{sec:softcore-model}

The function $f$ above should be chosen based on the phenomenon we would like to model. The activated sweat gland location patterns are repulsive. Our first attempt was to use a hardcore model, where sweat glands cannot be closer together than some minimum hardcore distance, but such a model turned out not to be flexible enough. Therefore, we suggest to use a
soft-core model with the density 
\begin{align*}
f_{SC}(y; \vec{\bf x}_k,  R, \kappa) \propto \exp\left(-\sum_{i=1}^{k}\left(\frac{R}{d(y, x_i)}\right)^{2/\kappa}\right)
\end{align*}
for adding the point $y$ in the realisation. Above, $R>0$ is an inhibition range parameter and $0<\kappa<1$ in the exponent describes how "soft-core" the model is. In the limit as $\kappa\rightarrow 0$, we obtain a hard-core process with hard-core distance $R$. Some  soft-core Gibbs point process models have been introduced in the literature \citep{OgataTanemura1981,OgataTanemura1984}, including models with the particular interaction function that we use here
\citep{spatstat2015}.

The log likelihood of the model becomes 
\begin{equation}\label{seqmodel_loglik}
l(R, \kappa; \vec{\bf x}_n) 
= -\log|W| -\sum_{k=2}^n\sum_{i=1}^{k-1}\left(\frac{R}{d(x_k, x_i)}\right)^{2/\kappa} - \sum_{k=2}^n\log Z(R, \kappa, \vec{\bf x}_k),
\end{equation}
where
\begin{equation*}
    Z(R, \kappa, \vec{\bf x}_k)^{-1} = \int_W \exp\left(-\sum_{i=1}^{k-1}\left(\frac{R}{d(y, x_i)}\right)^{2/\kappa}\right) dy
\end{equation*}
is a normalising constant.

\subsection{Efficient likelihood inference for the sequential models}\label{sec:seqmodel_inf}

Even though the likelihood of a sequential point process can be costly to compute, the particular sum structure in  \eqref{seqmodel_loglik} allows faster computations. Using an integration scheme with $J$ integration points $y_1, y_2, \dots, y_J$ with weights $w_1, w_2, \dots, w_J$, the last term in \eqref{seqmodel_loglik} can be written as
\begin{align*}
    \sum_{k=2}^n\log Z(R, \kappa, \vec{\bf x}_k)
    &= \sum_{k=2}^n\log \left(\int_W \exp\left(-\sum_{i=1}^{k-1}\left(\frac{R}{d(y, x_i)}\right)^{2/\kappa}\right) dy\right)^{-1}\\
    &= - \sum_{k=2}^n\log \sum_{j=1}^J w_j\exp\left(-\sum_{i=1}^{k-1}\left(\frac{R}{d(y_j, x_i)}\right)^{2/\kappa}\right).
\end{align*}
In total, there are $Jn(n-1)/2$ summands, among which only $Jn$ are distinct. Therefore, the integrals are efficiently calculated by evaluating the terms in the innermost sum only once.

\subsection{Soft-core model with noise}\label{sec:mixture-model}

To account for the incorrectly identified close pairs in the extracted point patterns, we used a mixture model where one of the components is a uniformly distributed error component. Such an error component can be added to any point process model and here, we add it in the sequential soft-core model.
The arrival density of a point $y$ (after the uniformly distributed first point) is then
\begin{align*}
f_M(y; \vec{\bf x}_k, R, \kappa, \theta) &= (1-\theta) f_{SC}(y; \vec{\bf x}_k, R, \kappa) + \frac{\theta}{|W|}\\
&= (1-\theta) Z(R, \kappa, \vec{\bf{x}}_k)\exp\left(-\sum_{i=1}^{k}\left(\frac{R}{d(y, x_i)}\right)^{2/\kappa}\right) + \frac{\theta}{|W|}.
\end{align*}
Therefore, the point at $y$ comes from the soft-core process with probability $1-\theta$ (the first term on the right-hand side of the formula) and from the uniformly distributed error process with probability $\theta$. Even though this model allows extra points everywhere, not only near the real activated glands, it can improve estimation of the  parameters as shown below. However, the parameter $\theta$ cannot be interpreted directly as the probability of incorrectly identified glands since some of the points without close neighbours regarded as noise could as well be true glands.

The log-likelihood of the soft-core model with uniformly distributed error is given by
\begin{equation}\label{seqmodelnoise_loglik}
    l_M(R, \kappa, \theta; \vec{\bf x}_n) = -\log |W| + \sum_{k=2}^n \log f_M(x_k; \vec{\bf x}_{k-1}, R, \kappa, \theta).
\end{equation}

\subsection{Application to the sweat gland data}\label{sec:seqmodel_data}

The soft-core model was fitted without and with noise to each sweat gland point pattern independently. First, we compared the maximum likelihood estimates of the soft-core parameters obtained without or with added noise. Then, we fitted the model with noise to the data in a Bayesian framework to be able to better compare the goodness-of fit of the sequential soft-core model and the generative model presented in Section \ref{sec:genmodel}. 
We used regular grid based integration with 10\,800 integration points to evaluate the likelihood in all cases.

\subsubsection{Parameter estimates without and with added noise}

\begin{figure}
\centering
\includegraphics[width=\textwidth]{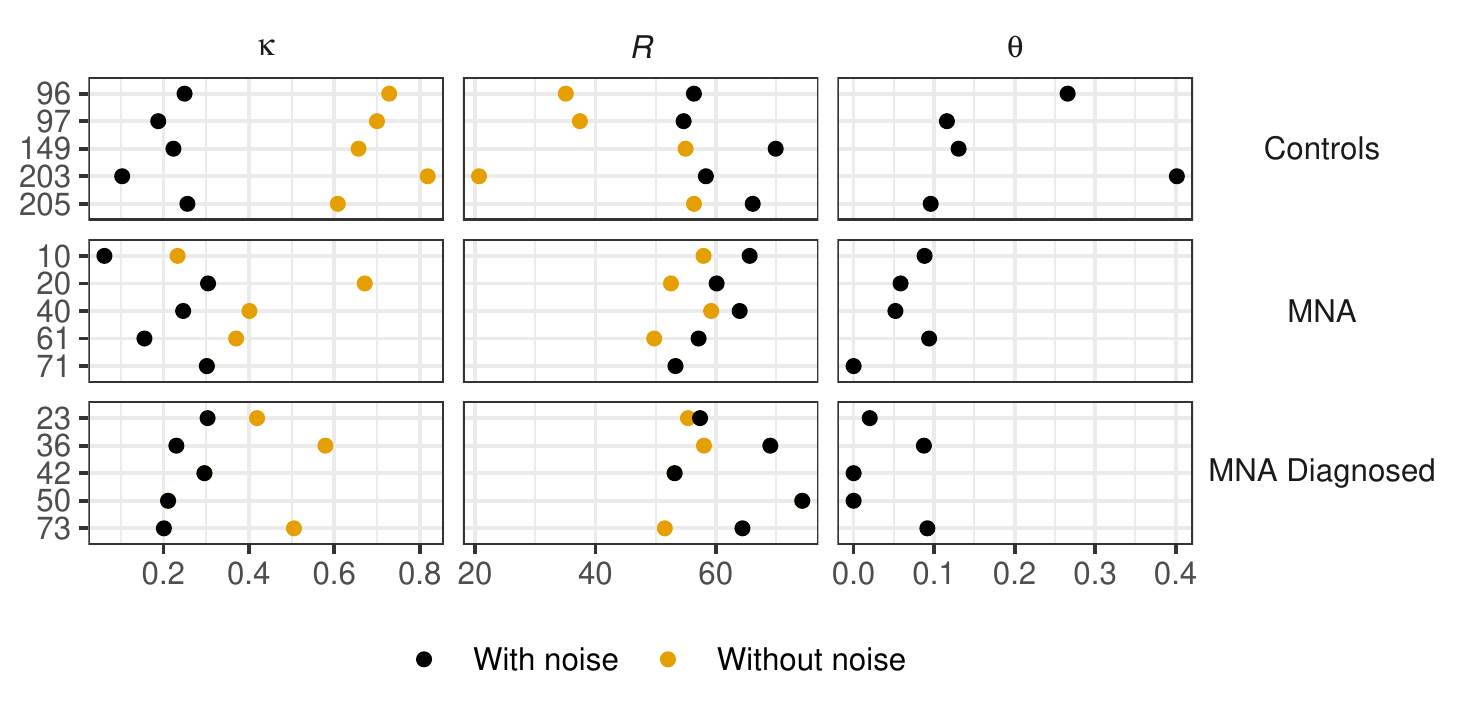}
\caption{Parameter estimates of the softcore model without (Softcore) and with (Mixture) noise fitted separately to each subject of the three groups (subject numbers shown on the left).}
\label{fig:parestSoftcore}
\end{figure}

The parameter estimates obtained by maximizing the log likelihood \eqref{seqmodel_loglik} or \eqref{seqmodelnoise_loglik} with respect to the parameters can be seen in Figure \ref{fig:parestSoftcore}, where black dots belong to the sequential soft-core model without noise and the yellow dots to the model with noise.
The estimates obtained without noise for the range parameter $R$ are on average smaller and the "softness" parameter $\kappa$ larger in the control group than in the neuropathy groups. However, for the model fitted with noise, only the mixture parameter $\theta$, which is estimated larger for the control group than for the neuropathy groups, differs between the groups. 

\begin{figure}
\centering
\includegraphics[width=\textwidth]{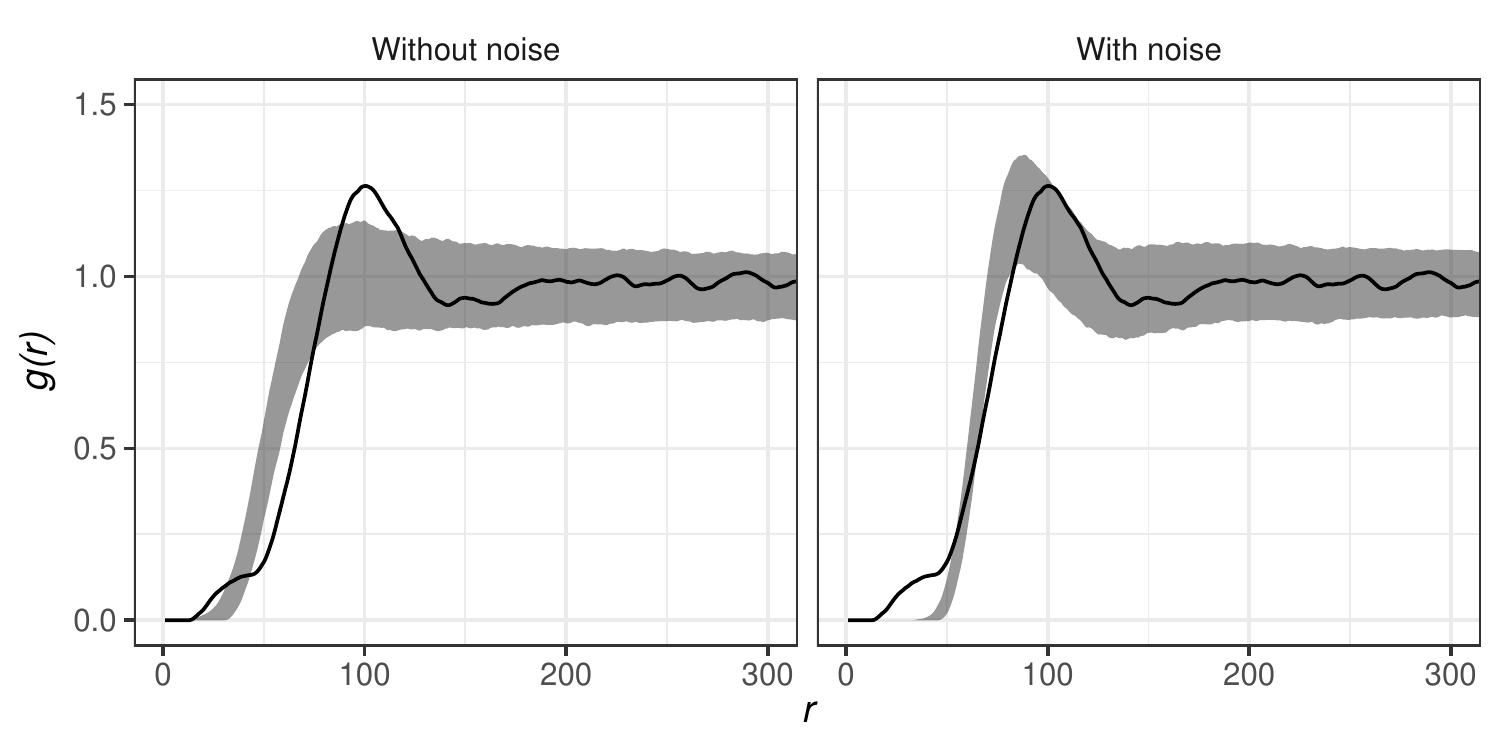}
\caption{Empirical pair correlation functions (black lines) for subject 205 in the end of the video recording together with 95\% global envelopes (grey areas) constructed from 25000 simulations from the soft core model estimated without (left) and with (right) noise.}
\label{fig:envelopesNoise}
\end{figure}

We investigated the goodness-of-fit of the fitted softcore models by using the pair-correlation function. We generated samples from the sequential soft-core models with parameters $R$ and $\kappa$ estimated with and without noise. The uniform noise was not simulated.
Figure \ref{fig:envelopesNoise} shows the empirical pair-correlation functions for subject 205 for the softcore model estimated with and without noise together with 95\% global envelopes \citep{MyllymakiEtal2017, MyllymakiMrkvicka2019} calculated from 25000 samples of each model. 
It can be seen that for this subject, the range parameter is clearly underestimated if estimation is done without accounting for noise. For the other subjects, the goodness-of-fit of the model with noise was also as good or better than the goodness-of-fit of the model without noise. The bad fit of the model at short distances is explained by the incorrectly recorded close pairs of points that are present in the data but not in the simulations.

\subsubsection{Bayesian inference of the model with noise}
    
\begin{figure}
\centering
\includegraphics[width=\textwidth]{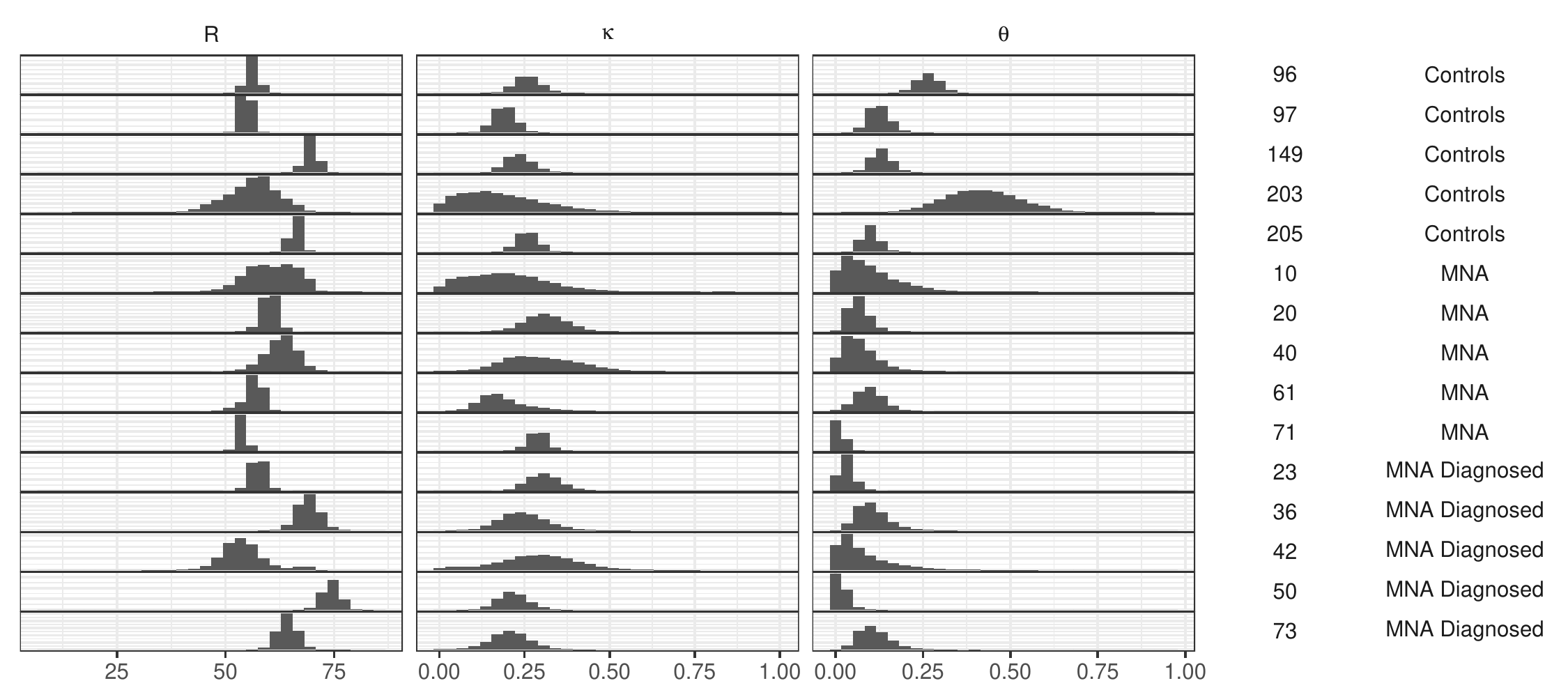}
\caption{Posterior marginals for each subject (rows) and each parameter (columns) for the soft core model estimated with noise.}
\label{fig:parestSoftcoreBayes}
\end{figure}

We fitted the soft-core model with noise to the sweat gland data also by using standard likelihood-based Bayesian approach with Robust Adaptive Metropolis algorithm \citep{Vihola2012}. We ran the MCMC for 120\,000 iterations and discarded the first 20\,000 iterations as burn-in.
As the prior distribution for the range parameter $R$ we used Gamma distribution with shape parameter $3$ and scale parameter $70/3$

and the prior for $\kappa$ and $\theta$ was the uniform distribution on $[0,1]$.
The posterior histograms in Figure \ref{fig:parestSoftcoreBayes} show some variation within the groups but no clear differences between the groups: The arrival density parameters $R$ and $\kappa$ were estimated to be rather similar in all groups. The $\theta$ parameter related to the errors appears to be somewhat larger in the control group than in the other two groups.

\begin{figure}
\centering
\includegraphics[width=\textwidth]{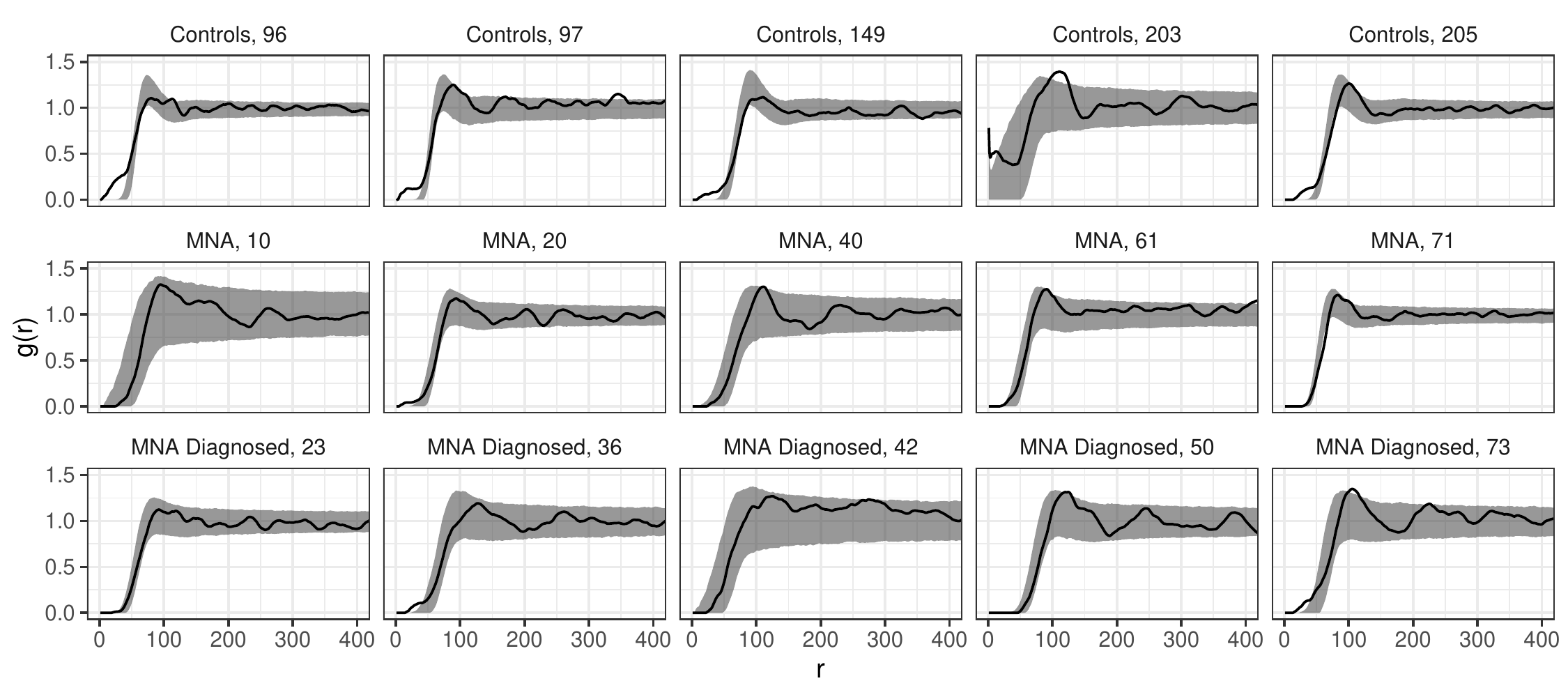}
\caption{Empirical pair correlation functions (black lines) for each subject in the end of the video recording together with 95\% global envelopes (grey areas) constructed from 25000 simulations from the posterior predictive distribution of the soft core model estimated with noise.}
\label{fig:envelopesSoftcoreBayes}
\end{figure}

Figure \ref{fig:envelopesSoftcoreBayes} shows the empirical pair-correlation functions for each subject together with the global envelopes \citep{MyllymakiEtal2017, MrkvickaEtal2018} calculated from 25000 simulations from the posterior predictive distribution of the fitted softcore models with noise.
In most cases, the envelopes cover the empirical curves. For some subjects, especially for the controls, the empirical pair-correlation function is not covered by the envelopes at very short distances. This is expected, as mentioned earlier, since according to the model used, this behaviour is caused mainly by noise, which was not simulated. The envelopes are quite wide close to the peak of the curves and do not always capture the shape of the peak particularly for the patients who have smaller number of activated sweat glands. To conclude, we did not find any differences in the arrival density of the sweat glands between the groups based on the fifteen studied subjects.

\section{Generative point process model}\label{sec:genmodel}

In our second approach, we first model the underlying unobserved sweat glands and then, model the activated sweat glands as an independent thinning of the underlying gland pattern. 
Modelling the glands and the activation of them separately allows one to answer questions regarding specifically the activation process. 
One possible hypothesis is that the underlying gland pattern itself is not different between controls and subjects with neuropathy, but the activation process is different. 
More specifically, almost all glands should activate on healthy subjects while the glands of the subjects with neuropathy could have a tendency to leave larger holes in the activation process \citep{ProviteraEtal2010}.

\subsection{Model specification}

It seems reasonable to assume that the underlying (unobserved) sweat gland pattern is a rather densely packed regular point pattern covering the whole skin. To obtain such a structure, some type of soft-core sequential inhibition process, where points are added as long as it is possible (we do not know the actual number of glands), would be appropriate. However, it is not straightforward to decide when to stop adding points since theoretically, soft-core type of interaction always allows new points. Instead, we start by generating a simple sequential inhibition (SSI) model\citep{IllianEtal2008}, which is then disturbed to obtain a soft-core structure.
A sample from the SSI model is generated sequentially by proposing points from the uniform distribution and accepting them if the pattern satisfies the hardcore condition with hardcore distance $R$, i.e.\ the new proposed point does not lie within distance $R$ from any earlier point. This is continued until there is no space left for new points. 
The disturbed SSI model is obtained from the "pure" SSI model by displacing the location of each point with an independent zero mean isotropic Gaussian random variable with covariance $\sigma^2 I$.

We assume independent gland activation, i.e.\ that the final pattern is a result of an independent thinning of the underlying disturbed SSI process. Therefore, the model has three parameters: inhibition range $R$, hardness of inhibition $\sigma$, and probability of activation $p$.

\subsection{Parameter estimation using approximate Bayesian computation}

For the generative model, we cannot write down the likelihood. However, if we make the simplification that the generation of SSI points is continued until, for example, 300 failed attempts are proposed in a row, sampling from the model is easy. Approximate Bayesian computation (ABC) is a method for Bayesian inference in situations where the likelihood of the model is intractable\citep{MarinPudloRobertRyder2012, SunnokerEtal2013}, but it is possible to simulate the model.
It is based on sampling from the (pseudo-) posterior distribution
\begin{align*}
 \pi_\epsilon(\theta) = \pi(\theta)\P(\| s(Y_{\theta}) - s(y) \| < \epsilon),
\end{align*}
where $Y_{\theta}$ follows the model with parameter vector $\theta$, $y$ is the data, $\pi(\cdot)$ the prior distribution for the parameters, $s$ an appropriately chosen summary statistic, and $\epsilon$ a tolerance level. 

\subsubsection{ABC-MCMC}

A simple ABC rejection sampler is expressed in the following algorithm

\begin{algorithm}
\caption{A simple ABC rejection sampler}
\begin{algorithmic}
 \For{$i \gets 1,M$}
  \Repeat
   \State Generate parameter vector $\theta'$ from the prior distribution $\pi$
   \State Generate a realisation $z$ from the model with parameter vector $\theta'$
  \Until{$\|s(z) - s(y)\| \le \epsilon$}
  \State $\theta_i \gets \theta'$
 \EndFor
\end{algorithmic}
\end{algorithm}

\noindent This basic algorithm can be rather inefficient, but fortunately, there are several more efficient algorithms for performing ABC.
We used an adaptive ABC-MCMC algorithm \citep{ViholaFranks2019}.
In our data study below, the MCMC was run for $10\,000\,000$ iterations and the $250\,000$ simulated parameter values with the smallest distances $\|s(z) - s(y)\|$ were taken as the posterior sample.

\subsubsection{Summary statistics}

The choice of summary statistics is crucial for the ABC method to work. For a regular point process model, it is natural to use summary statistics based on the pair correlation function $g$. Instead of using the full pair correlation function $g$, we tried to find a specific part of it that would be sufficient for our purpose following the rule of thumb\citep{LiFearnhead2018} that the number of summary statistics in the ABC approach should approximately match with the number of parameters to be estimated. The location of the first peak of the pair correlation function is intuitively connected to the inhibition range $R$. However, the location of the first peak can be difficult to estimate exactly and thus, we used the smallest distance $r_1 > 10$ pixels where $g(r_1) = 0.75$ as the location of the uphill before the first peak.
Furthermore, the slope of the uphill provides information on the "softness" parameter $\sigma$ and we chose the smallest distance $r_2 > 10$ pixels where $g(r_2) = 1$ as the second summary statistic.
Finally, the smallest distance $r_3$ in the empty space function $F$ where $F(r_3) = 0.5$ was taken as the third summary statistic to represent the activation probability $p$.
The empty space function was chosen because it gives information on the number of points but is not greatly affected by erroneous nearby points.
Since all the chosen summary statistics, $r_1, r_2$ and $r_3$, have a similar order of magnitude, we did not have to add any weights in the ABC algorithm.
The specific values 0.75 and 1 were chosen to be somewhat separated and not too small to account for possible errors caused by splitting of spots into multiple glands that would cause the pair-correlation function not to start from zero. In addition, we only considered distances greater than 10 pixels since at very short distances the kernel estimator of the pair correlation function is not very reliable. These choices worked well for the sweat gland data, as demonstrated below.

\subsection{Application to the sweat gland data}

The generative model was fitted to the sweat gland data using the ABC approach described above. 
In addition to the above specifications, we needed to set the priors.
For $R$ we used improper, uniform prior on $[40, \infty)$ restricting that $R$ could not be unreasonably small, while in addition to being unrealistic, small $R$ values result in a large number of points in the SSI process which is computationally challenging. The prior of $p$ was uniform on $[0.1, 1]$, stating that at least 10\% of the glands (modelled by the underlying disturbed SSI process) needed to activate and thus be observed. Furthermore, for $\sigma$ we used the gamma distribution with the shape parameter equal to $10/3$ and scale parameter equal to $3$. 
While the priors $R$ and $p$ can be considered rather non-informative, the prior for $\sigma$ was somewhat informative suggesting positive, but not too large $\sigma$. Note that if $\sigma$ was very large in comparison to $R$ it would break all the structure of the SSI process, which is unreasonable.

\begin{figure}
\centering
\includegraphics[width=\textwidth]{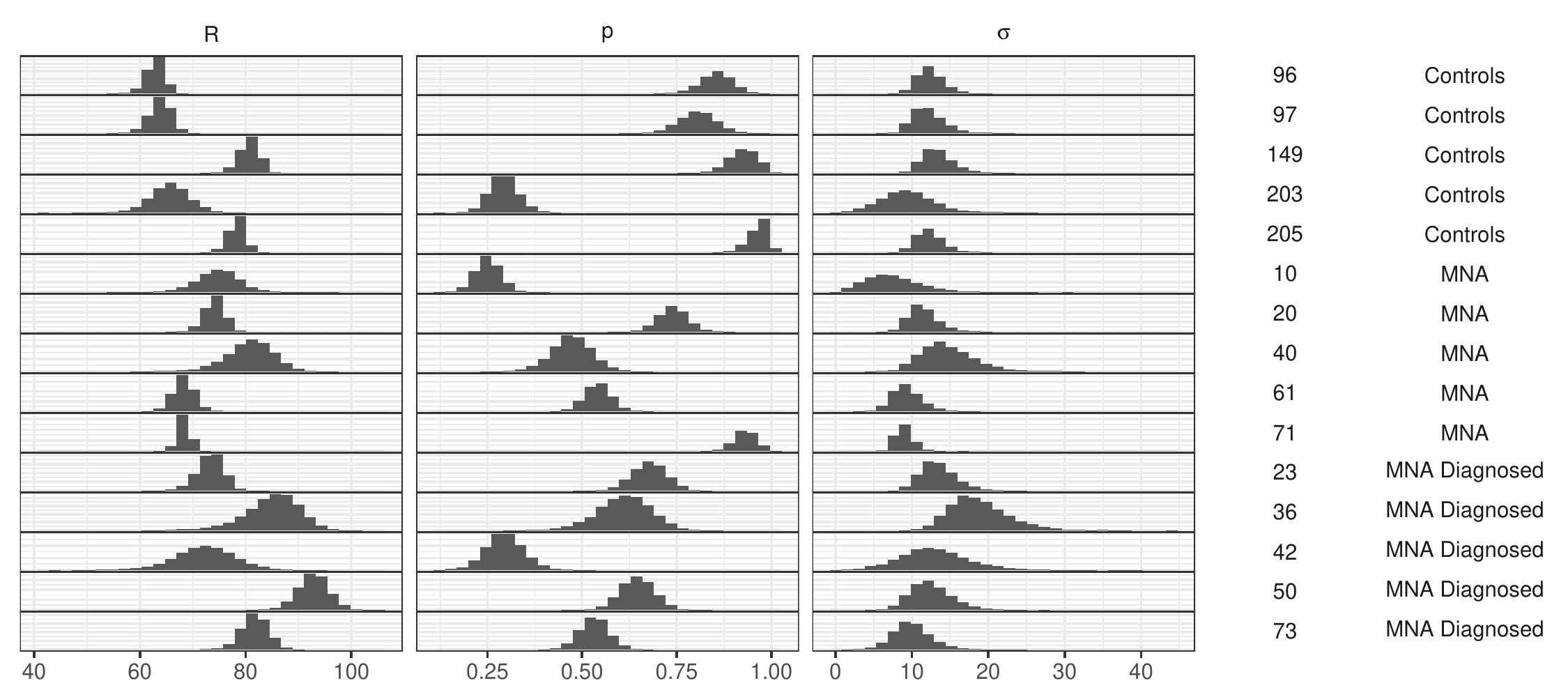}
\caption{Posterior marginals for each subject (rows) and each parameter (columns) for the generative model. }
\label{fig:parestABC}
\end{figure}

\begin{figure}
\centering
\includegraphics[width=\textwidth]{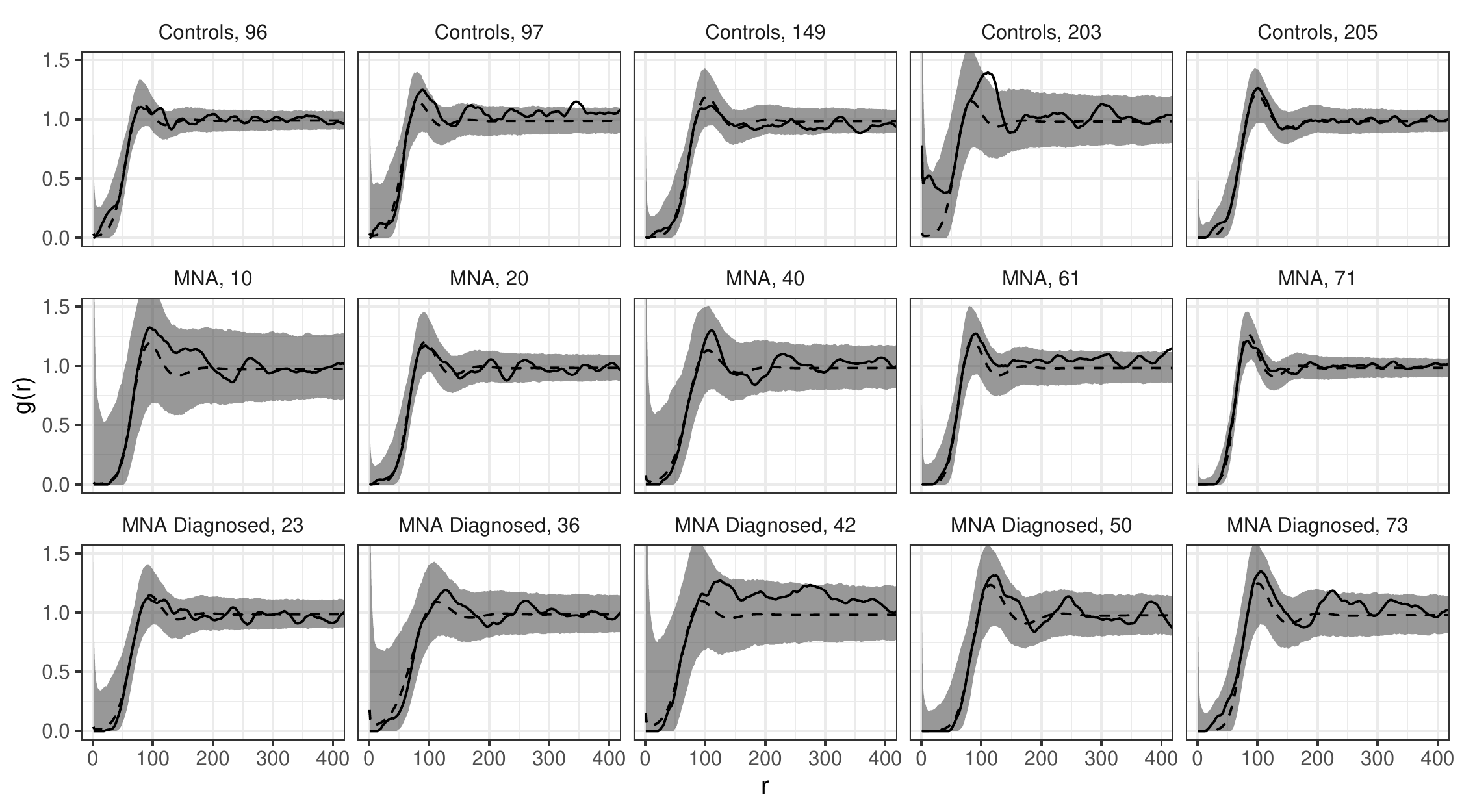}
\caption{Empirical pair correlation functions (black lines) for each subject in the end of the video recording together with 95\% global envelopes (grey areas) and means (dashed lines) constructed from 25000 simulations from the posterior predictive distribution of the generative model.}
\label{fig:envelopesSSIABC}
\end{figure}

The posterior marginal histograms for the parameters can be seen in Figure \ref{fig:parestABC} and 95\% global envelopes for the pair correlation function constructed from 25000 simulations from the posterior predictive distribution in Figure \ref{fig:envelopesSSIABC}.
As can be seen in Figure \ref{fig:parestABC}, the parameter estimates vary somewhat between the subjects and groups. Differences in the softness of the model,\ i.e.\ in the values of the parameter $\sigma$, are small.
However, there seems to be a slight tendency for the inhibition range $R$ to be a little smaller in the control group than in the MNA groups, but the difference is not clear based on the limited amount of data we have. The range was always between 60 pixels and 100 pixels.
Furthermore, the control subjects tend to have a larger activation probability than the MNA patients, but the within group variation is large. This is in agreement with earlier studies, which indicate that a larger number of sweat glands of controls than of MNA patients activate\citep{LoavenbruckEtal2017}.

\begin{figure}
\centering
\includegraphics[width=\textwidth]{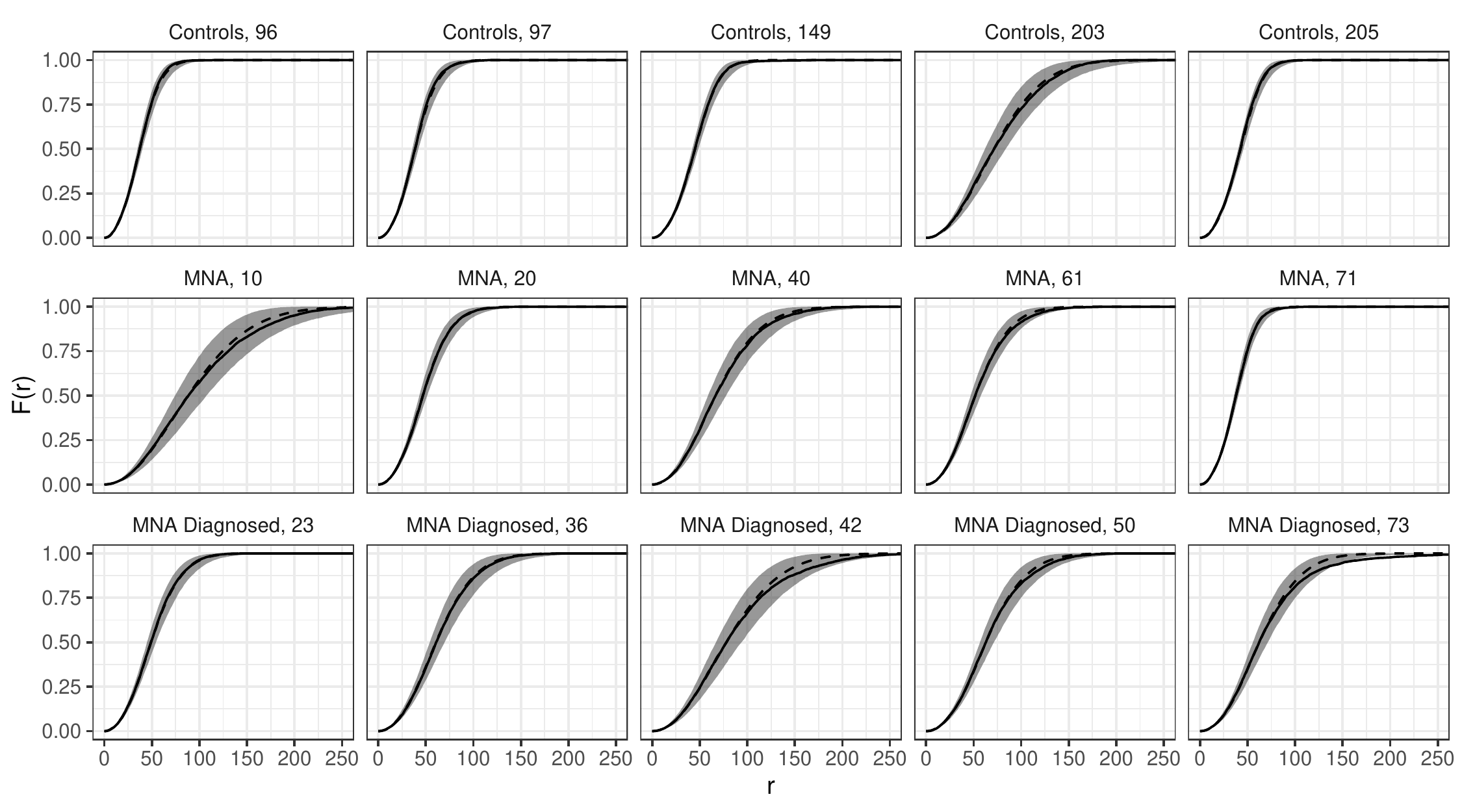}
\caption{Empirical empty space functions (black lines) for each subject in the end of the video recording together with 95\% global envelopes (grey areas) and means (dashed lines) constructed from 25000 simulations from the posterior predictive distribution of the generative model. }
\label{fig:envelopesSSIABCF}
\end{figure}

According to the visual evaluation of the global envelopes of the pair-correlation function (see Figure \ref{fig:envelopesSSIABC}) and empty space function (see Figure \ref{fig:envelopesSSIABCF}),
the model seems to fit quite well to the data. It captures the behavior of the pair correlation function both at small distances and around the first top.
It should also be mentioned that the envelopes for the pair correlation function are rather wide at small distances covering the observed functions almost in all cases, even though the model did not include any error term. The wide envelopes are due to the relatively wide posterior distribution of $\sigma$. Namely, large $\sigma$ can lead to some close pairs in the patterns and consequently also positive values of the pair correlation function at small distances. Another reason for the relatively wide envelopes may be that the summary statistics used in the ABC approach were chosen such that they do not use any information at very short distances. 

We explored a few other priors for $\sigma$, namely improper uniform and exponential distributions with means $1, 2$ and $4$. The posterior distributions of the other parameters were not affected by the choice of the prior for $\sigma$, but the posterior of $\sigma$ itself was somewhat sensitive to the choice and also the goodness-of-fit of the model measured by the pair correlation was affected. Namely, improper uniform prior led to wider posterior distribution of $\sigma$ and large $\sigma$ caused the variation of the pair correlation function to be even higher at small distances. On the other hand, the strict exponential priors shrank the posterior distribution towards zero, and very small $\sigma$ caused the peak of the pair correlation function to be too sharp. Thus, the disturbance parameter $\sigma$ needed a somewhat informative prior to lead to a good fit of the model.

\begin{figure}
\centering
\includegraphics[width=\textwidth]{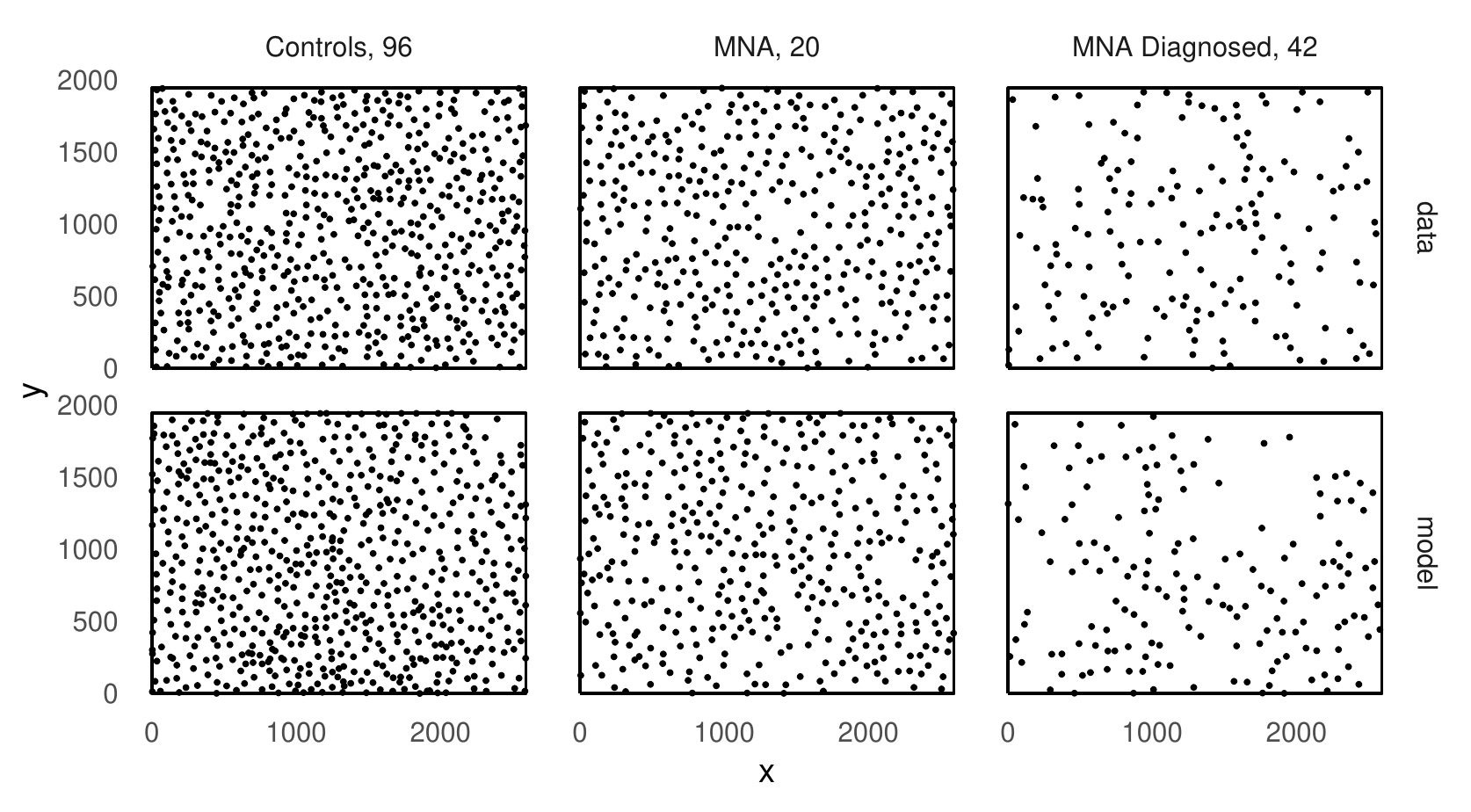}
\caption{The original point patterns (top) and patterns generated from the corresponding posterior predictive distributions of the generative model (bottom) for three subjects.}
\label{fig:pointpatternexamples}
\end{figure}

We simulated patterns from the posterior predictive distribution and the simulated patterns mimic the data patterns rather well, see Figure \ref{fig:pointpatternexamples}. Note, in particular, that the independent thinning seems to produce rather similar empty spots as there are in the data, as also indicated by the empty space function (Figure \ref{fig:envelopesSSIABCF}).

\section{Discussion}\label{sec:discussion}

We suggested two point process models for the activation of sweat glands, a sequential softcore model describing the appearance of the activated sweat glands and a thinned disturbed SSI process, that we call a generative model, where we start by modelling the underlying unobserved sweat gland pattern. Data were videos of sweat gland activation recorded from 15 subjects. As one of the image analysis steps needed to extract the locations of the sweat glands, we proposed a change point detection approach to decide whether a pixel belongs to a wet area. For automatic selection of the threshold for detecting a change point, we investigated the possibility to use a simple statistical model (not presented). However, choosing the thresholds manually turned out to give better results. A manual choice of thresholds in fact gave a reasonable way to adapt to some pecularities in the videos such as darkening in time due to other reasons than sweat.

Maximizing the log-likelihood function of a sequential point process has been regarded as computationally costly due to the integrals in the normalizing constants\citep{PenttinenYlitalo2016}. 
However, for the sequential softcore model these integrals have a particular sum form which allows to compute them efficiently and to perform Bayesian inference. The same efficient computation scheme is applicable for any sequential point process having an arrival density with a similar sum structure. To estimate the parameters of the generative model, we employed an ABC algorithm since the likelihood function was not easily available. 

Even though our proposed image analysis approach worked well, there were some incorrectly identified close pairs of glands in the extracted point patterns. To take into account such errors, we added an error term  in the sequential softcore model resulting in a mixture model having a softcore component and a uniform noise component. For the generative model, on the other hand, the summary statistics in the ABC approach were chosen such that they were robust to close pairs of points. 

The proposed models were fitted to the data and the parameters estimated from the patterns from healthy subjects and from subjects suffering from neuropathy were compared. 
The activation probability (in the generative model) was higher in the control group than in the neuropathy groups. Based on the limited amount of data, we were not able to find any further differences between the groups.
Our generative model with independent activation fitted the data well. 
It also seems that the independent activation can result in similar holes in the point patterns as observed earlier  in the sweat gland patterns\citep{ProviteraEtal2010}, see e.g.\ the bottom right plot in Figure \ref{fig:pointpatternexamples}. 

We believe that the models suggested here, especially the generative model, are good starting points for further studies using larger data sets including more subjects and replicates from each subject. It would certainly be interesting to include the sizes of the sweat spots into the analysis and explore whether the independent activation is adequate even when more data and information are available.

\section*{Acknowledgements}

Mikko Kuronen and Mari Myllym{\"a}ki have been financially supported by the Academy of Finland (Project Numbers 306875 and 295100) and Aila Särkkä by the Swedish Research Council (VR 2013-5212) and by the Swedish Foundation for Strategic Research (SSF AM13-0066). The authors thank Matti Vihola for useful discussions.

\bibliography{}

\end{document}